\documentclass[prb,aps,twocolumn,superscriptaddress,showpacs]{revtex4-1}

\usepackage{graphicx}
\usepackage{epstopdf}

\begin{document}

\title{Helical spin-waves, magnetic order, and fluctuations in the langasite compound Ba$_{3}$NbFe$_{3}$Si$_{2}$O$_{14}$}

\author{C. Stock}
\affiliation{NIST Center for Neutron Research, 100 Bureau Drive, Gaithersburg, Maryland 20899, USA}
\affiliation{Indiana University Cyclotron Facility, 2401 Milo B. Sampson Lane, Bloomington, Indiana 47404, USA}

\author{L.C. Chapon}
\affiliation{ISIS Facility, Rutherford Appleton Laboratory, Chilton, Didcot, OX11 0QX, United Kingdom}

\author{A. Schneidewind}
\affiliation{Forschungsneutronenquell Heinz Meier-Leibnitz (FRM-II), D-85747 Garching, Germany}
\affiliation{Institut fur Festkorperphysiki, TU Dresden, D-1062 Dresden, Germany}

\author{Y. Su}
\affiliation{Juelich Centre for Neutron Science, IFF, Forschungszentrum Juelich, Outstation at FRM II, Lichtenbergstr. 1, D-85747 Garching, Germany}

\author{P.G. Radaelli}
\affiliation{Oxford Physics, Clarendon Laboratory, Parks Road, Oxford OX1 3PU, United Kindom}

\author{D.F. McMorrow}
\affiliation{London Centre for Nanotechnology, Department of Physics and Astronomy, University College London, London WC1E 6BT, United Kingdom}

\author{A. Bombardi}
\affiliation{Diamond Light Source Ltd., Harwell Science and Innovation Campus, Didcot, Oxfordshire OX11 0DE, United Kingdom}

\author{N. Lee}
\affiliation{Rutgers Center for Emergent Materials and Department of Physics and Astronomy, Rutgers University, 136 Frelinghuysen Rd., Piscataway, New Jersey 08854, USA}

\author{S.-W. Cheong}
\affiliation{Rutgers Center for Emergent Materials and Department of Physics and Astronomy, Rutgers University, 136 Frelinghuysen Rd., Piscataway, New Jersey 08854, USA}

\date{\today}

\begin{abstract}

We have investigated the spin fluctuations in the langasite compound Ba$_{3}$NbFe$_{3}$Si$_{2}$O$_{14}$ in both the ordered state and as a function of temperature.  The low temperature magnetic structure is defined by a spiral phase characterized by magnetic Bragg peaks at $\vec{q}$=(0,0,$\tau\sim1/7$) onset at T$_{N}$=27 K as previously reported in Ref. \onlinecite{Marty08:101}.  The nature of the fluctuations and temperature dependence of the order parameter is consistent with a classical second order phase transition for a two dimensional triangular antiferromagnet.  We will show that the physical properties and energy scales including the ordering wavevector, Curie-Weiss temperature, and the spin-waves can be explained through the use of only symmetric exchange constants without the need for the Dzyaloshinskii-Moriya interaction.  This is accomplished through a set of ``helical" exchange pathways along the $c$ direction imposed by the chiral crystal structure and naturally explains the magnetic diffuse scattering which displays a strong vector chirality up to high temperatures well above the ordering temperature.  This illustrates a strong coupling between magnetic and crystalline chirality in this compound.

\end{abstract}


\maketitle

\section{Introduction}

	In geometrically frustrated magnets, each spin cannot satisfy all pair wise interactions as a result of the crystal symmetry and therefore remain disordered to temperatures well below the Curie-Weiss temperature ($\Theta_{CW}$) where magnetic order is expected.~\cite{Collins97:75,Ramirez01:13,Kawamura01:79}   The simplest geometrically frustrated system is arguably the two dimensional triangular lattice where all interactions cannot be satisfied resulting in large degeneracies.~\cite{Kawamura98:10} 

	Low-spin (or quantum) triangular antiferromagnets with $S\leq 1$ do not show long-range magnetic order because of the combination of strong quantum fluctuations and geometrical frustration and a number of interesting phases have been observed and proposed including resonating valence bond states and spin liquids.~\cite{Anderson73:8}   Examples of two-dimensional (2D) triangular antiferromagnets include the Cs$_{2}$CuCl$_{4}$ (S=1/2), $\kappa$-(BEDT-TTF)$_{2}$Cu$_{2}$(CN)$_{3}$ (S=1/2), ZnCu$_{3}$(OH)$_{6}$Cl$_{2}$ (Kagome) and NiGa$_{2}$S$_{4}$ (S=1).~\cite{Coldea03:68,Nakatsuji05:309,Shimizu03:91,Helton07:98,Lee07:6} All of these systems display anomalous magnetic properties well below $\Theta_{CW}$ and illustrate the dramatic effects that crystal symmetry imposed degeneracy, or geometric frustration, can have on the magnetic structures and excitations.

	Large $S$ systems are interesting to investigate as quantum fluctuations are completely suppressed and therefore reveal only the classical effects of frustration.  An interesting example is the case of the anisotropic triangular lattice $\alpha$-NaMnO$_{2}$ where the $S=2$ spins, with dominant nearest neighbor coupling, relieve the effects of frustration by reducing the dimensionality and behave like one-dimensional chains.~\cite{Stock09:103}  More complex systems include (Tb,Ho)MnO$_{3}$ (Refs. \onlinecite{Kenzelmann05:95,Vajk05:94,Senff07:98}) and RbFe(MoO$_{4}$) (Ref. \onlinecite{Kenzelmann07:98}) where frustration results in spiral phases and coupling between magnetic and ferroelectric order parameters.~\cite{Kimura03:426} For some of these systems, the loss of inversion symmetry in the ordered magnetic state allows antisymmetric exchange (Dzyaloshinskii-Moriya interaction) which has been proposed to be very important to multiferroicity.  It is therefore important to investigate the property of chirality and its relationship to the crystal structure especially in systems that are non-centrosymmetric already in the paramagnetic state.~\cite{Kawamura01:79}

	An interesting class of compounds is the langasite series of materials which consist of isolated triangular lattices (or trimers) on a hexagonal lattice.  Attention to this system started with La$_{3}$Ga$_{5}$SiO$_{14}$ (LGS) which displays favorable piezoelectric properties and strong electromechanical coupling constants.~\cite{Kitaura04:69}  LGS has a Ca$_{3}$Ga$_{2}$Ge$_{4}$O$_{14}$ type structure with space group P321 and this structure type can accomodate a number of different cations leading to a wide variety of different properties.  For example, Pr$_{3}$Ga$_{5}$SiO$_{14}$ has drawn some attention as a two-dimensional Kagome system which only magnetically orders under chemical pressure.~\cite{Zhou09:102,Lumata10:81}  The only strongly magnetic langasite compounds discovered so far are those containing iron ions and have been observed to display a series of novel piezoelectric and magnetic properties.~\cite{Marty10:81}

\begin{figure}
\includegraphics[scale=0.4]{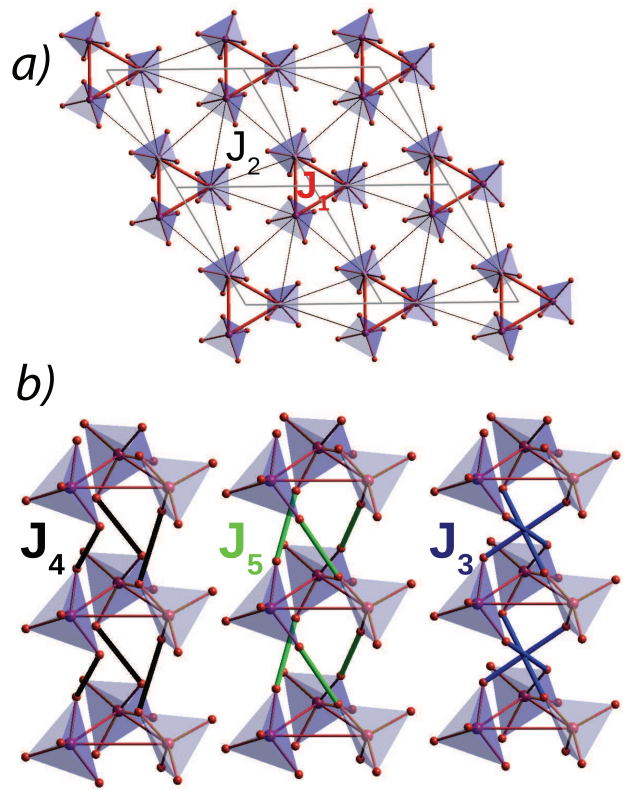}
\caption{(Color online) Crystal structure of Ba$_3$NbFe$_3$Si$_2$O$_{14}$ projected in the $ab$-plane (top panel) and along the $c$-axis (bottom panel). The FeO$_4$ tetrahedra are shown in light blue color. The Fe and O atoms are represented as large and small spheres respectively. The Ba,Nb and Si atoms are omitted for clarity. Exchange interactions are shown in plane by thick line (J$_1$) and thin lines (J$_2$). Along the $c$-axis, the three inequivalent exchange interactions are marked by their respective O-O paths, shown as thick bonds.}
\label{structure_figure}
\end{figure}

	A particular iron based langasite compound is Ba$_{3}$NbFe$_{3}$Si$_{2}$O$_{14}$ which consists of Fe$^{3+}$ $S=5/2$ ions.  A detailed structural study of this compound has been presented previously illustrating a unique single phase magnetic spiral structure with each triangle adopting a conventional 120$^{\circ}$ arrangement but with a spiral pitch along the $c$ axis resulting in a magnetic propagation vector of $q_{0}$=(0,0,$1/7$).~\cite{Marty08:101}  The underlying structure is displayed in Fig. \ref{structure_figure}.  Panel $a)$ illustrates the Fe$^{3+}$ framework in the $a-b$ plane and shows the material consists of triangles of magnetic ions arranged on a hexagonal lattice.  The lower panel $b)$ illustrates the framework along the $c$ axis. The space group is chiral and the crystal possess a handedness, as seen from the ``helical" paths along the $c$-axis (panel b). The exchange pathways labeled J$_{i=1-5}$ are discussed later in this paper.  The magnetic order occurs at T$_{N}$=27 K despite a Curie-Weiss temperature of $\Theta_{W}$=-173 K, owing to the effects of low-dimensionality and frustration. Further work on this material has suggested the possibility of multiferroic properties at low temperature and as a function of applied magnetic field.~\cite{Marty10:81,Zhou09:21}  

	We present a neutron elastic and inelastic scattering study of the fluctuations and critical properties in Ba$_{3}$NbFe$_{3}$Si$_{2}$O$_{14}$. We show that the magnetic chirality is imposed by symmetric Heisenberg exchange only without the need for a dominant Dzyaloshinskii-Moriya interaction and illustrates a strong coupling between structural and magnetic chirality. The paper is divided into four sections discussing our results along with an introduction and a section on the experimental details.  We first investigate the temperature dependence of the magnetic correlations near the critical wave vector through the use of inelastic powder and single crystal studies to show that the transition behaves classically as expected for a two dimensional triangular magnet. This section also includes our results using polarized neutrons to investigate the energy integrated diffuse scattering as a function of temperature.  To motivate our inelastic work, we then present our results from spin-dimer calculations which provide an estimate for the relative values of the magnetic exchange parameters.  We then discuss the spin-waves and our heuristic model to extract the exchange constants.  We finish the paper with a derivation of the ordering wave vector and Curie-Weiss temperature.  While Dzyaloshinskii-Moriya interactions are present owing to the symmetry of the lattice, we find that we can understand the ordering wave vector (q$_{0}$=(0,0,$\sim$1/7)), the spin-waves, and Curie Weiss constant ($\Theta_{W}$) in terms of a model based solely on Heisenberg exchange with a magnetic chirality introduced because of the handedness forced by the crystal structure.  

\section{Experimental Details}

	Neutron experiments utilized instruments at both the ISIS spallation neutron source (Rutherford Appleton Laboratory, UK) and the FRM2 reactor (Germany). To initially characterize the dynamics and temperature dependence, we used the MARI direct geometry chopper spectrometer at the ISIS facility. The sample consisted of 11 g of powder mounted in an annular geometry and cooled using a closed cycle displex.  A Gd Fermi chopper was used to define an incident energy in parallel with a disc chopper to suppress background and high-energy neutrons above $\sim$ 250 meV.  Fast neutrons, with energies in excess of $\sim$ 1 eV, were removed using a nimonic (or $t_{0}$) chopper located close to the target face.  To measure the magnetic fluctuations as a function of energy and temperature we used a fixed incident energy of E$_{i}$=10 meV taken with a Fermi chopper speed of 250 Hz. 

	The single crystal sample, with mass 6 g, was grown using the floating zone technique and was characterized at the test and alignment spectrometer at the ISIS facility using a four-circle geometry. The lattice constants were measured to be $a$=$b$= 8.539 \AA, and $c$=5.241 \AA\ with $\gamma$=120$^{\circ}$.  The sample was aligned such that Bragg reflections of the form (H0L) lay within the horizontal scattering plane.  Two different sets of experiments were performed at the FRM2 reactor using a closed cycle refrigerator to reach temperatures as low as 2.5 K.  To measure the spin-waves across the entire Brillouin zone, the PANDA cold triple-axis was used.  A vertically focused pyrolytic graphite (PG002) monochromator was used to select an incident neutron energy reflected onto the sample position.  A flat PG002 analyzer was used to select the final energy of the neutrons analyzed in the detector.  A cooled Beryllium filter on the scattered side filtered out higher order neutrons reflected off the monochromator and collimation sequence was set to open-80$'$-$S$-80$'$-open.  Final energies of E$_{f}$=5.0 and 2.5 meV were chosen to study the low-energy spin waves and E$_{f}$=13.5 meV was used to study the magnetic order parameter with the Beryllium filter replaced by a graphite filter.  All data taken on PANDA have been corrected for higher order contamination of the incident beam monitor which determines the counting time.  The correction factor is discussed in detail elsewhere.~\cite{Shirane_book,Stock04:69}

	To measure the energy integrated diffuse magnetic scattering, we have used the DNS polarized cold 2 axis diffractometer located at FRM2.  An incident energy of 3.64 meV was selected using a horizontally and vertically focused PG002 monochromator.  The beam was polarized using $m=3$ Scharpf supermirror polarizers.  The polarization at the sample was fixed through the use of an $xyz$ coil with the $x$ direction chosen to be parallel to the average $\vec{Q}$ at the sample position and $z$ vertical.  With the use of flipping coils in the incident and scattering beams, the two spin-flip and non spin flip cross sections could be measured with the neutron polarization along the three orthogonal Cartesian coordinates (a total of 12 channels).  The flipping ratio was 20 $\pm$ 1 and was not found to deviate from this value regardless of the direction of neutron polarization.  All spin-flip data has been corrected for the feed through from the non-spin-flip channel.  The scattered beam was measured with 24 detectors equally spaced 5$^{\circ}$ apart covering a total angular range in scattering angle ($2\theta$) equal to 120$^{\circ}$.  These measurements were performed using a two-axis geometry and therefore provided an approximate measure of the energy integrated intensity ($S(\vec{Q})$).

\section{Paramagnetic scattering and magnetic order}

\subsection{Magnetic order and wave vector}

\begin{figure}[t]
\includegraphics[width=90mm]{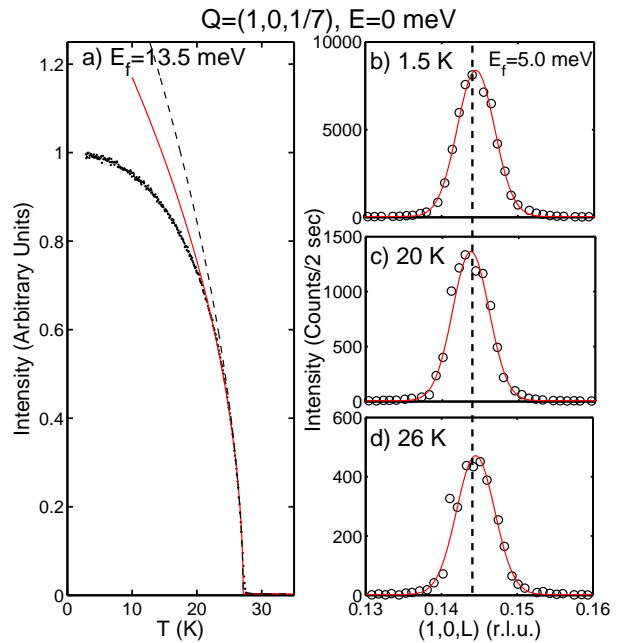}
\caption{(Color online) $a)$ The temperature dependence of the elastic magnetic scattering measured with a final energy of E$_{f}$=13.5 meV taken on the PANDA cold triple-axis.  The curves are power laws descrbed in the text.  Panels $b)-d)$ are scans along the (00L) direction illustrating that the incommensurability remains fixed with the temperature.} 
\label{elastic_figure}
\end{figure}

	A summary of the elastic scattering which characterizes the magnetic structure is presented in Fig. \ref{elastic_figure} and illustrates the temperature dependence of the magnetic Bragg peak at $\vec{Q}$=(1,0,$\sim$1/7).  The temperature dependence of the magnetic Bragg peak is plotted in panel $a)$ and was taken using the PANDA cold triple-axis with E$_{f}$=13.5 meV. The solid red curve is a plot to a power law $I\propto(T_{c}-T)^{2\beta}$ with T$_{c}$=27 K and $\beta$=0.25 following the critical exponents measured in CsMnBr$_{3}$, and CsVCl$_{3}$.~\cite{Kawamura88:63}  All of these systems have demonstrated to be XY-like in nature with a 120$^{\circ}$ structure similar to that in one trimer in Ba$_{3}$NbFe$_{3}$Si$_{2}$O$_{14}$.

	Panels $b)-d)$ in Fig. \ref{elastic_figure} illustrate scans along the (00L) direction at several temperatures.  The results illustrate that while there is a significant change in the intensity, the wave vector of the magnetic peak does not change within error even close to the Neel transition temperature.  This point is unusual with chiral magnets or incommensurate systems as the chirality (and hence wave vector) can be considered as a measure of the order parameter and this should vary smoothly near T$_{N}$ unless the transition is first-order.  For example, the field driven commensurate-incommensurate transition in CuGeO$_{3}$ is first order and becomes second order on doping with non magnetic impurities such as Mg resulting in a continuous change of the incommensurate wave vector near T$_{N}$.~\cite{Christianson02:66}  Another example is the incommensurate transition in Rb$_{2}$ZnCl$_{4}$ which exhibits a strong temperature dependence of the wave vector near the critical temperature.~\cite{Zinkin96:54}  To track the incommensurability as a function of temperature we now investigate the energy integrated intensity which is characteristic of the critical spin fluctuations using polarized neutrons.

\subsection{Polarized diffuse scattering measurements}

\begin{figure}[t]
\includegraphics[width=90mm]{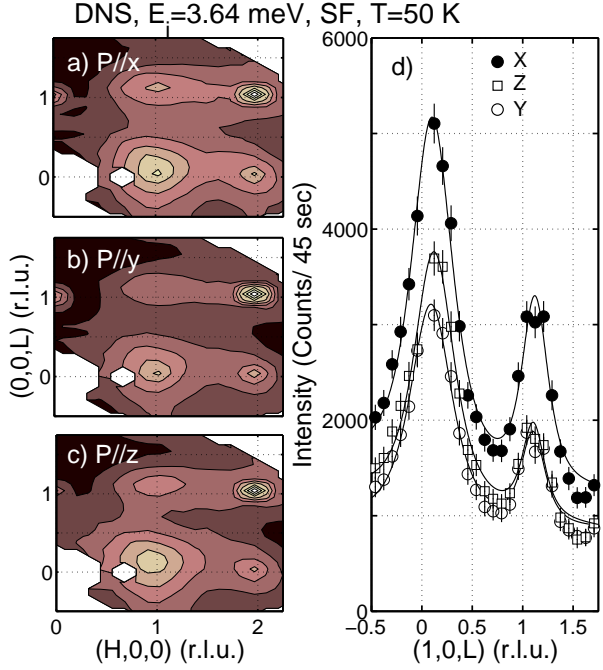}
\caption{(Color online) The three spin-flip cross sections measured with the incident beam polarized parallel to $x$, $y$, and $z$.  A cut along the L direction is displayed in panel d) for all three cross sections.  The cut was integrated over the range $0.9<H<1.1$.} 
\label{figure_iso}
\end{figure}

\begin{figure}[t]
\includegraphics[width=75mm]{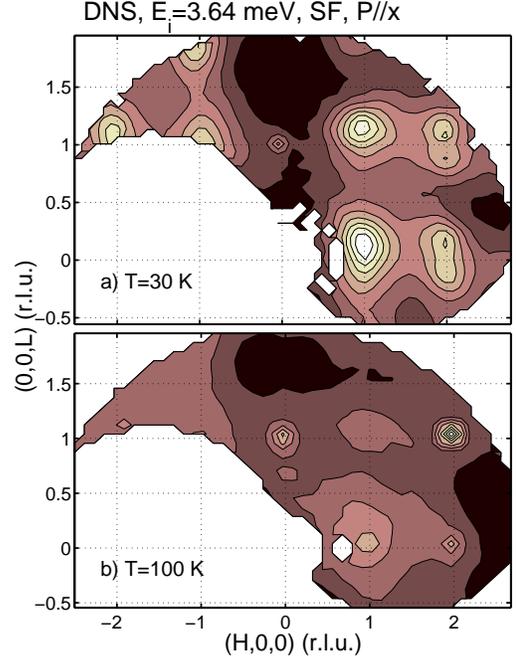}
\caption{(Color online) The spin flip cross section with the incident polarization parallel to $x$ is illustrated at $a)$ 30 K and $b)$ 100 K.  The data are plotted on a log scale.  The data were taken on DNS with an E$_{i}$=3.64 meV.} 
\label{pol_general_figure}
\end{figure}

	The spin of the neutron allows a sensitive measure of both the \textit{phase and magnitude} of the magnetic cross section.  The cross sections for polarized neutrons have been studied theoretically and are presented in Refs. \onlinecite{Halpern39:55,Blume63:130,Moon69:181,Schweika10:xx}.  Of particular interest is the spin-flip (SF) cross section when the incident beam is polarized parallel to the momentum transfer ($\vec{Q}$),

\begin{eqnarray}
\label{pol_formula}
I^{+-/-+}=\sum\limits_{ij} e^{\vec{Q}\cdot(\vec{r}_{i}-\vec{r}_{j})}p_{i}p_{j}^{*}[\vec{S}_{i\perp}\cdot\vec{S}_{j\perp}\mp ...\nonumber \\ i\hat{z}\cdot(\vec{S}_{i\perp}\times\vec{S}_{j\perp})].
\end{eqnarray}

\noindent Here $p$ is the magnetic scattering length and $S_{\perp}$ is the value of the magnetic spin perpendicular to the momentum transfer.  The symbols $+-/-+$ denote the cross sections when scattering from a spin-up to a spin-down state and from a spin-down to a spin-up respectively.  In collinear magnets, the values of these cross sections are the same as the second term in the above equation is identically zero.  

	In a chiral magnet, one can define a vector chirality via a vector product of two neighboring spins, averaged over three spin pairs by 

\begin{eqnarray}
\label{chiral_formula}
\vec{\kappa}={{2}\over{3\sqrt{3}}}\sum_{i,j} [\vec{S}_{i} \times \vec{S}_{j}],
\end{eqnarray} 
	
\noindent which is related to the cross product in the cross section written in Eqn. \ref{pol_formula}.  In a chiral magnet the cross product in Eqn. \ref{chiral_formula} is not necessarily equal to zero and I$^{+-}$ is not equal to I$^{-+}$.  This property of chiral magnets and polarized neutron scattering was first demonstrated in MnSi where the Dyzaloshinski-Moriya interaction results in spiral magnetic correlations.~\cite{Shirane83:28} Given that in our notation $x$ is parallel to the average $\vec{Q}$, a subtraction of scattering channels in $+x$ and $-x$ channels gives the cross product term above, sensitive to the chirality. The cross product in Eqn. \ref{pol_formula} does not appear when the incident beam polarization is perpendicular to the momentum transfer ($\vec{Q}$).  This is the case of for $y$ and $z$ polarizations and these two channels can be used to determine the magnetic moment direction or the orientation of $\vec{S}_{\perp}$ in the above formula.

	Fig. \ref{figure_iso} provides an overview of the magnetic scattering cross section in the paramagnetic phase by displaying the spin-flip intensities for polarizations parallel to the three orthogonal directions $x$, $y$, and $z$ at T=50 K.  A cut through the data along the (1,0,L) direction is plotted in panel $d)$ for the three polarizations.  The spin-flip cross section parallel to $x$ is a measure of the entire magnetic cross section. As illustrated in panel $d)$, the spin flip cross section parallel to $x$ is approximately twice that of the $y$ and $z$ channels.  This indicates that the scattering is isotropic to a first approximation, as expected for paramagnetic scattering above T$_{N}$.  There is a difference between the spin-flip cross sections in the $y$ and $z$ channels  with the $z$ cross section being larger than the $y$ channel at $\vec{Q}$=(1,0,0).  This indicates a small anisotropy between the paramagnetic scattering above T$_{N}$ and maybe expected given the highly anisotropic structure reported.~\cite{Marty08:101}  

\begin{figure}[t]
\includegraphics[width=90mm]{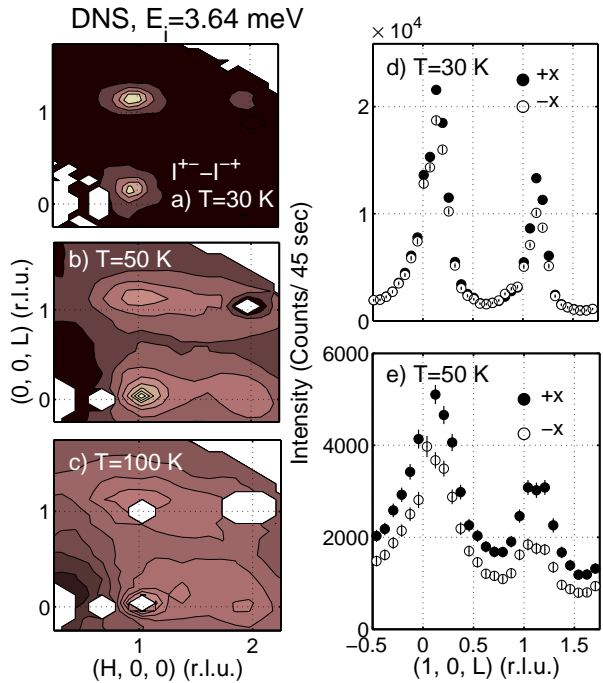}
\caption{(Color online) The spin flip cross section with the incident polarization parallel to $x$ and $-x$.  Panels $a)-c)$ illustrate the difference between the spin flip cross sections $I^{+-}$ and $I^{-+}$ at several temperatures.  Panel $b)$ and $c)$ are linear cut along (1,0,L) illustrating the difference between the two cross sections. The cuts was integrated over the range $0.9<H<1.1$.} 
\label{figure_SF_x}
\end{figure}

	Fig. \ref{pol_general_figure} illustrates the temperature dependence of the contours of magnetic scattering above T$_{N}$ at 30 K (panel $a)$) and 100 K (panel $b)$).  Both data sets were taken with the incident beam polarization parallel to $x$ and therefore are a measure of the total magnetic scattering cross section.  The data are plotted on a logarithmic scale.  Since $x$ is parallel to the average $\vec{Q}$ and no analyzer crystal was used, these scans provide an approximate measure of the total energy integrated magnetic scattering cross section.  At T=30 K, well defined diffuse scattering contours exist around the incommensurate positions at $q_{0}$=(0,0,$\sim$ 0.15) where magnetic ordering was observed below T$_{N}$.  At T=100 K, clear diffuse scattering is present though much reduced in intensity and considerably broader in momentum indicating a decrease in the correlation length.  It is interesting to note that the contours of constant scattering intensity imply an anisotropic line shape which is broader in the [100] direction than the [001].  This implies that the c-axis correlations are significantly stronger and therefore it is important to understand the c-axis magnetic exchanges in understanding the critical and physical properties of this system.  We will return to this point later when discussing the spin-waves at low-temperatures.  

	As pointed out in Ref. \onlinecite{Kawamura88:63}, a vector chirality variable can be defined by Eqn. \ref{chiral_formula} and is proportional to the cross product in the second term of Eqn. \ref{pol_formula}.  This chirality should occur simultaneously with spin ordering and follow a power law divergence similar to the magnetic order parameters in conventional second order phase transitions.  This order parameter can be derived in polarized neutron scattering experiments as the difference between the two spin-flip cross sections $I^{+-}$ equal to $I^{-+}$ when the neutron polarization is parallel to the momentum transfer vector.  Therefore, near the phase transition the difference between the two spin-flip channels and the value of the incommensurability should to go zero.  We have investigated this quantity using the DNS polarized neutron diffractometer.  

\begin{figure}[t]
\includegraphics[width=90mm]{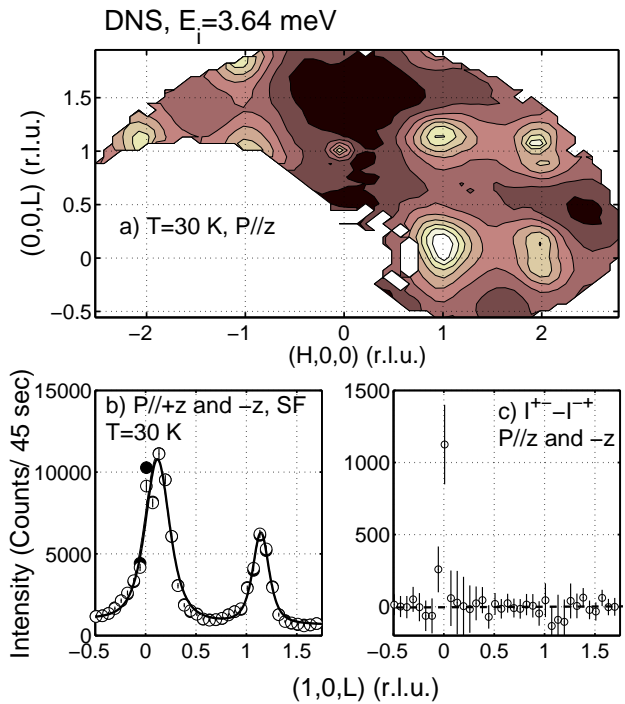}
\caption{(Color online) The spin flip cross section with the incident polarization parallel to $z$ and $-z$.  Panel $a)$ illustrates contours of the raw intensity with beam parallel to $z$ and hence shows $I^{+-}$.  Panel $b)$ illustrates a comparison between the $I^{+-}$ and $I{-+}$.  The cuts was integrated over the range $0.9<H<1.1$. Panel $c)$ is a subtraction of the two cross sections illustrating no difference within experimental error.} 
\label{figure_SF_z}
\end{figure}

	The difference between the SF cross sections $I^{+-}$ and $I^{-+}$ with the incident neutron polarization parallel to the $x$ axis is displayed in Fig. \ref{figure_SF_x}.  Panels $a-c)$ plot subtracted intensity $I^{+-}-I^{-+}$ and is sensitive to the chirality parameter discussed above.  If there were no chirality present, then the cross product in the Eqn. \ref{pol_formula} would vanish and the difference would be zero.  All panels at temperatures ranging from 30-100 K clearly show the presence of correlated magnetic scattering with a vector chiral component, even at 100 K well above the ordering temperature of 27 K.  The magnetic scattering is also clearly displaced away from the nuclear Bragg peak positions and located close to the position in momentum where the magnetic Bragg peak is observed at low temperatures.  As these measurements integrate in energy, we are not able to determine from these scans if this diffuse scattering is dynamic or static or on what time scale the fluctuations occur at.  This will be discussed in the next section with energy resolved triple-axis and direct geometry chopper measurements.  The white gaps near the nuclear Bragg peak positions in Fig. \ref{figure_SF_x} are regions where the strong feed through from the non spin flip channel as a result of nuclear scattering did not subtract well and hence have been removed from the data for clarity.  

	To confirm the experimental setup, we plot a similar analysis but with the incident polarization parallel to the $z$ axis in Fig. \ref{figure_SF_z}.  This is an important check as the difference observed above could result from differences in the flipping ratios between the $+x$ and $-x$ configurations.  One of the two cross sections ($I^{+-}$) is illustrated in panel $a)$ with panel $b)$ plotting a cut of both $I^{+-}$ and $I^{-+}$.  Panel $c)$ confirms that there is no difference between the two cross sections to within experimental error except near the Bragg position (L=0) where the feed through from the non spin-flip cross section contaminates the data and does not subtract out fully.  This complete subtraction is the result expected for the spin-flip cross section when the incident beam polarization is perpendicular to the momentum transfer.  This confirms the quality of the experiment, the results, and the conclusions applied to the data taken with the polarization parallel to $x$ described above.

	The results here are quite surprising given that the chirality (as defined in Ref. \onlinecite{Kawamura88:63}) is predicted to diverge at T$_{N}$, yet we still observe a strong chirality at temperatures more than three times T$_{N}$.  This is in contrast to the results reported in MnSi which is a prototypical spiral magnetic resulting from the Dzyaloshinski-Moriya interaction.  In that work, it was found that the difference between the two cross sections became significantly smaller at high temperatures above the ordering temperature.  Indeed a Landau free-energy expansion did show that the two cross sections ($I^{+-}$ and $I^{-+}$) should become equal at high temperatures, consistent with the experimental data.  This is clearly not the case in Ba$_{3}$NbFe$_{3}$Si$_{2}$O$_{14}$ and implies that a picture utilizing Dzyaloshinski-Moriya must be examined carefully.  In the next section we discuss the energy resolved critical dynamics measured with both a chopper instrument and a cold triple-axis.
	
\subsection{Spin fluctuations and critical properties}

\begin{figure}[t]
\includegraphics[width=85mm]{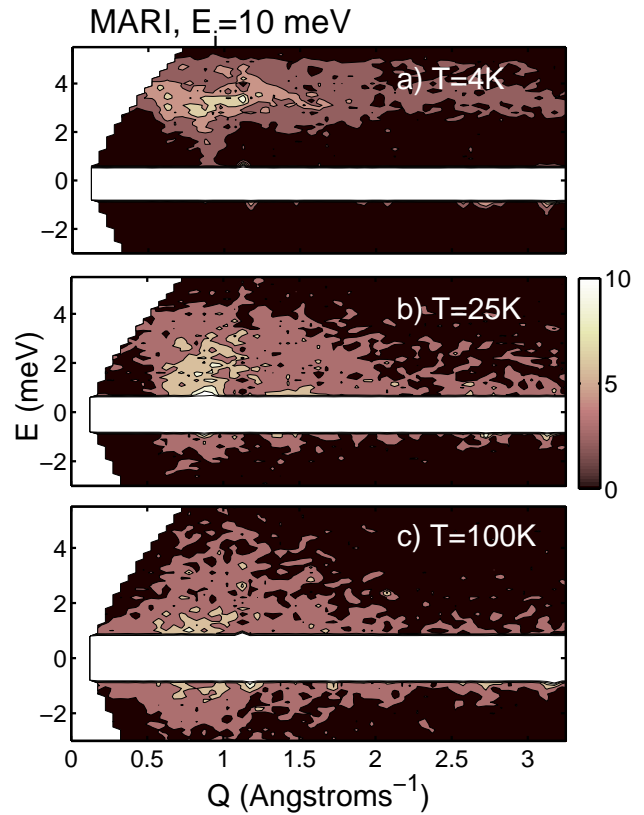}
\caption{(Color online) A series of contour plots of the magnetic scattering in Ba$_{3}$NbFe$_{3}$Si$_{2}$O$_{14}$ at a series of temperatures.  The data were taken on the MARI direct geometry chopper spectrometer located at ISIS with a fixed incident energy E$_{i}$=10 meV and at a Fermi chopper speed of 250 Hz.} 
\label{mari_inelastic_figure}
\end{figure}

\begin{figure}[t]
\includegraphics[width=70mm]{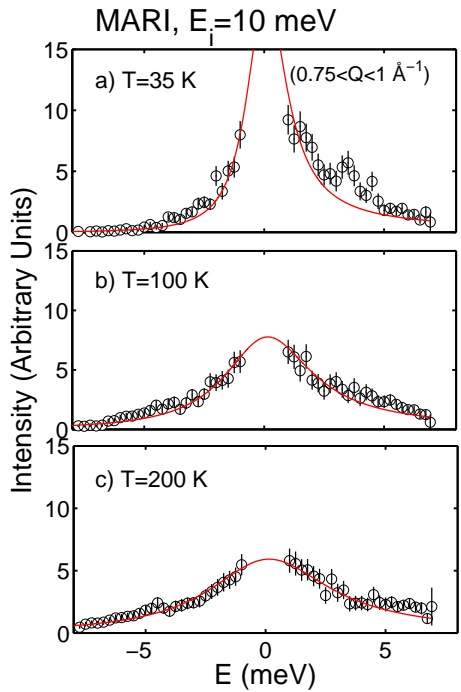}
\caption{(Color online) A series of constant-$Q$ cuts taken from the E$_{i}$=10 meV MARI data above T$_{N}$.  The solid curves are fits to the modified Lorentzian discussed in the text.  The data illustrate strong critical fluctuations which broaden with increasing temperature.} 
\label{mari_constantQ}
\end{figure}

	To understand the critical properties, we investigated the critical spin fluctuations near the ordering temperature using both powder and single crystal samples.  The inelastic spectrum at several temperatures in the powder is summarized in Fig. \ref{mari_inelastic_figure} with the data taken on MARI using an E$_{i}$=10 meV.  The low-temperature (T=4 K) spectrum will be discussed in detail later but the higher temperatures illustrate the presence of strong magnetic fluctuations above the ordering temperature of T$_{N}$=27 K.  To extract a line width as a function of temperature we have fit a series of constant-$Q$ cuts to the following formula,   

\begin{eqnarray}
\label{mod_lor}
S(E)=\chi_{0} [n(E)+1] {{E\Gamma}\over{\Gamma^{2}+E^{2}}}
\end{eqnarray}

\noindent where $\chi_{0}$ is the real part of the susceptibility, $\Gamma$ is the energy half width, and $n(E)+1$ is the Bose factor.  This formula allows us from a constant-$Q$ scan to extract the real part of the susceptibility ($\chi_{0}$) and an energy line width ($\Gamma$).  Examples of the resolution convolved fits extracted from the MARI data above T$_{N}$ are shown in Fig. \ref{mari_constantQ}.  

\begin{figure}[t]
\includegraphics[width=70mm]{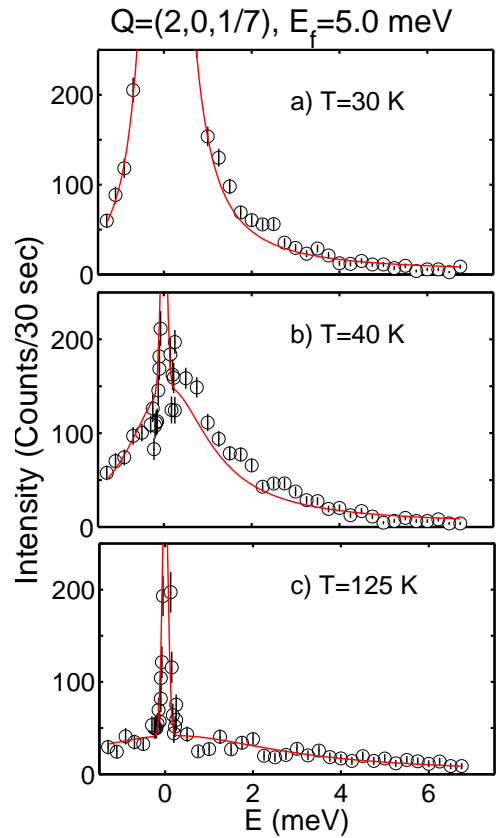}
\caption{(Color online) A series of constant-$Q$ scans taken on the PANDA cold triple-axis using a fixed final energy of E$_{f}$=5.0 meV at $\vec{Q}$=(1,0,1/7).  The solid curves are fits to the modified Lorentzian discussed in the text plus a the resolution function centered at the elastic energy (E=0 meV).} 
\label{panda_inelastic_T}
\end{figure}

	Constant-$Q$ cuts taken using the PANDA cold triple-axis spectrometer on single crystal samples are presented in Fig. \ref{panda_inelastic_T}.  These scans were performed at the measured ordering wave vector and the solid lines are fits to Eqn. \ref{mod_lor} plus a delta function centered at the $E=0$ elastic position defined by the energy resolution.  Similarly, from these scans we are able to extract parameters for the susceptibility $\chi_{0}$ and the energy line width $\Gamma$.

\begin{figure}[t]
\includegraphics[width=75mm]{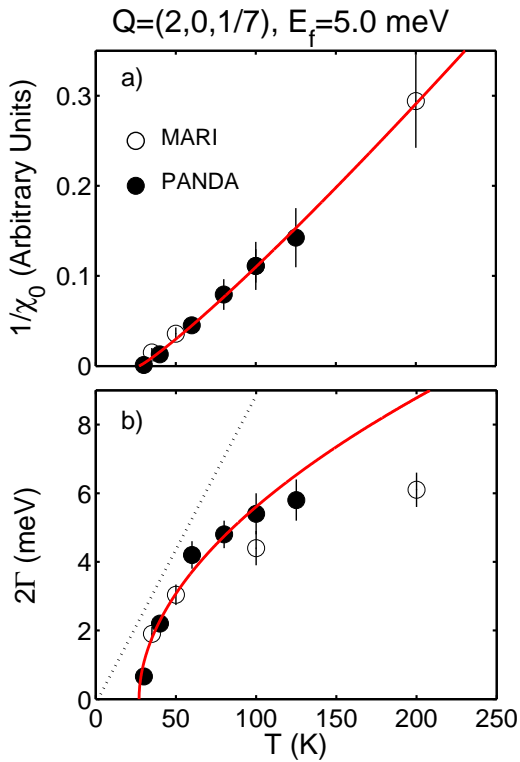}
\caption{(Color online) $a)$ A plot of the inverse of the real part of the susceptibility $1/\chi_{0}$ as a function of temperature for data taken on both powder (MARI) and single crystal (PANDA) samples. $b)$ The full-width ($2\Gamma$) is plotted against temperature.  The dashed line is $2\Gamma=k_{B}T$.  The solid curves in both panels are the power laws described in the text and previously measured in two dimensional triangular antiferromagnets. } 
\label{fit_param_inelastic}
\end{figure}

	A summary of the temperature dependence of $\chi_{0}$ and $\Gamma$ are illustrated in Fig. \ref{fit_param_inelastic}.  The scaling factor for the susceptibility between the MARI and PANDA data were obtained from a common temperature taken at 100 K.  The two data sets overlap very well illustrating a consistent analysis.  A plot of $\chi_{0}^{-1}$ (Fig. \ref{fit_param_inelastic} $a)$) illustrates a smooth function which intercepts the x-axis at the Neel temperature indicating that the transition is likely second order.  The solid curve is a plot of $\chi_{0}\propto (T-T_{N})^{-\gamma}$ with $\gamma$=1.1.  This is the same critical exponent predicted and measured for the strongly two dimensional triangular magnets CsMnBr$_{3}$, and CsVCl$_{3}$.~\cite{Kawamura88:63}  We consider these as good systems to compare the critical properties of Ba$_{3}$NbFe$_{3}$Si$_{2}$O$_{14}$ to as these triangular magnets display 120$^{\circ}$ magnetic ordering and weak coupling between triangular planes.  Also, because of strong dipolar interactions, these systems also display a strong XY character with the spins being forced in a particular plane. 

	A plot of the line width $2\Gamma$ as a function of temperature above T$_{N}$ is illustrated in panel $b)$ of Fig. \ref{fit_param_inelastic}.  The solid line is a plot to the power law $\Gamma\propto (T-T_{N})^{\nu}$ with $\nu$=0.53 as found for the two dimensional triangular systems described above.   The line width ($\Gamma$) can be related to the correlation length using hydrodynamic theory which states $\Gamma \propto \hbar c /\xi$, where $\hbar c$ is the spin-wave velocity, and $\xi$ is the correlation length.~\cite{Halperin69:188}  Given the correlation length scales as $\xi \propto (T-T_{N})^{-\nu}$ we have used the same power law form to describe the line width in panel $b)$ of Fig. \ref{fit_param_inelastic}.   

	The dashed line in Fig. \ref{fit_param_inelastic} $b)$ is a plot of $2\Gamma=k_{B}T$ and illustrates that the line width we observe is always less than the energy scale set by the temperature ($k_{B}T$).  This indicates that the dynamics and critical fluctuations are in the classical limit in contrast to frustrated magnets with lower spin-value, for example NiGa$_{2}$S$_{4}$ or in the superconductor YBa$_{2}$CuO$_{6+x}$, where $2\Gamma>k_{B}T$.~\cite{Nakatsuji05:309,Stock10:xx,Stock08:77}  These results indicate that the critical scattering in Ba$_{3}$NbFe$_{3}$Si$_{2}$O$_{14}$ is indeed classical in nature. 

	In summary, we have combined chopper and triple-axis measurements on single crystals and powders to obtain the temperature dependence of the energy linewidth ($2\Gamma$) and the susceptibility ($\chi_{0}$).  The temperature dependence is consistent with the power laws derived in other two-dimensional triangular  magnets where $\gamma=1.1$, $\beta=0.25$, and $\nu=0.53$ and indicates that the transition is second order and that the fluctuations near T$_{N}$ behave classically.~\cite{Kawamura88:63}  
	
\section{Spin dimer calculation}

	The crystal structure for the left-handed chirality is represented in Fig. \ref{structure_figure} projected in the ${ab}$-plane (top panel) and along the $c$-axis (bottom panel). Exchange paths are indicated by the label J$_i$ (i=1,5) following the conventions used in Ref. \onlinecite{Marty08:101}. All exchange terms are dominated by contributions from super super-exchange (SSE) through Fe$^{3+}$-O-O-Fe$^{3+}$ since there are no covalent Fe$^{3+}$-O-Fe$^{3+}$ bonds. J$_1$ and J$_2$ are respectively the nearest-neighbor and next-nearest neighbor exchange interactions in the ${ab}$-plane. Along $c$, there are three inequivalent exchanges paths connecting adjacent layers. J$_4$ connects each of the Fe$^{3+}$ ions in a triangular unit to ions directly above, i.e. related by the lattice translation (0,0,1). J$_3$ and J$_5$ links different Fe$^{3+}$ ions of adjacent triangles and create \emph{helical} paths along the $c$-axis of opposite chirality (anticlockwise for J$_3$ bonds and clockwise for J$_5$ bonds). The relative strengths of the different SSE terms are determined by the overlap of the O p wave functions, which highly depends on the value of the O-O distance with respect to the Van der Waals distance (2.80\AA) and the value of the Fe$^{3+}$-O-O-Fe$^{3+}$ dihedral angle. Relevant bond distances are presented in the upper row of Table \ref{table:spindimer}.  

\begin{figure}
\includegraphics[scale=0.35]{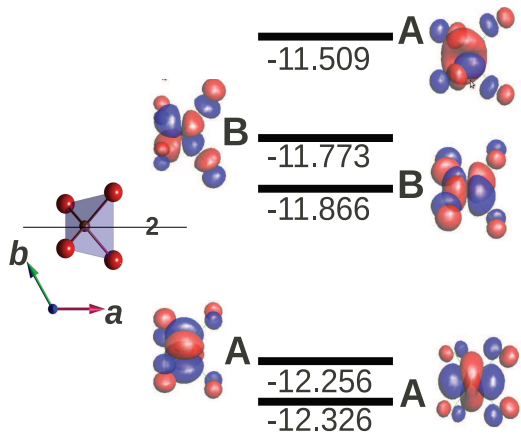}
\caption{(Color online) Molecular orbitals of the FeO$_4^{5-}$ complex calculated by the extended Huckel tight-binding method (see text for details). The five levels (all singlets) are shown with their corresponding energy in eV, symmetry label, and drawing of the molecular orbitals. On the left, a FeO$_4$ tetrahedra is shown in the same projection along the $c$-axis. The unit-cell axes, and position of the two-fold symmetry axis are also shown.}
\label{fig:MO}
\end{figure}

	The relative strengths of the five SSE interactions have been derived semi-quantitatively by a spin-dimer analysis based on extended Huckel tight-binding (EHTB) calculations.~\cite{Whangbo05:7} The atomic orbitals for Fe and O are approximated by double-$\zeta$ Slater orbitals, using the parameters listed in the supplementary information of Ref. \onlinecite{Koo06:45}. The Fe$^{3+}$ ion is in a distorted tetrahedral configuration (point symmetry 2), and the non-bonding E$_g$ levels and antibonding T$_{2g}$ levels found for Td symmetry splits into singlets E$_g\rightarrow$2A and T$_{2g}\rightarrow$A$\oplus$2B. The molecular orbitals of the FeO$_4^{5-}$ complex calculated by EHTB are shown in Fig. \ref{fig:MO}. Each the the singlet (labeled $\mu$=1,5 from the lowest to the highest energy) is occupied with a single electron due to the high-spin d$^5$ electronic configuration of Fe$^{3+}$. The antiferromagnetic contribution of each of the SSE interaction can be evaluated in turn by considering the corresponding Fe$_2$O$_8^{10-}$ dimer and calculating the average of the squared energy difference between the bonding and antibonding states:    

\begin{equation}
J\propto \langle\left(\Delta\epsilon\right)^2\rangle \propto \sum_{\mu=1,5} \left(\Delta\epsilon_{\mu\mu}\right)^2,
\label{dft_equation}
\end{equation}

\noindent where the sum runs over the five sets of bonding-antibonding states resulting from correlations between pairs of all the 3A and 2B molecular orbitals. Essentially the values of $<(\Delta\epsilon)^2>$ are proportional to the square of the hopping integral (t$^2$), i.e. to the exchange energies.The results are listed in Table \ref{table:spindimer}. The strongest exchange interaction is J$_1$, slightly stronger that the other in-plane exchange J$_2$. Along the $c$-axis, J$_5$ is by far the strongest exchange, as already pointed out in Ref. \onlinecite{Marty08:101}, and approximatively four times larger than J$_4$. Due to the very long O-O distance, J$_3$ is negligible in comparison with all other exchange terms.       

\begin{table}[h!]
\begin{tabular}{c|c|c|c|c|c}
 SSE interaction & J$_1$ & J$_2$ & J$_3$ & J$_4$ & J$_5$ \\
 \hline
 d(Fe-Fe) $\AA$ & 3.692 & 5.652 & 6.411 & 5.241 & 6.411 \\
 d(O-O) $\AA$ & 2.769 & 2.624 & 3.965 & 2.901 & 2.774 \\
 \hline
$\epsilon_{11}$ (eV) & -12.329 & -12.260 & -11.878 & -11.793 & -11.514 \\
$\epsilon_{11}*$ (eV) & -12.323 & -12.248 & -11.856 & -11.732 & -11.464 \\
\hline    
$\epsilon_{22}$ (eV) & -12.333 & -12.258 & -11.869 & -11.794 & -11.518 \\
$\epsilon_{22}*$ (eV) & -12.316 & -12.251 & -11.861 & -11.741 & -11.485 \\
\hline    
$\epsilon_{33}$ (eV) & -12.327 & -12.257 & -11.867 & -11.773 & -11.510 \\
$\epsilon_{33}*$ (eV) & -12.326 & -12.255 & -11.866 & -11.772 & -11.508 \\
\hline    
$\epsilon_{44}$ (eV) & -12.329 & -12.267 & -11.870 & -11.776 & -11.510 \\
$\epsilon_{44}*$ (eV) & -12.323 & -12.238 & -11.862 & -11.768 & -11.508 \\
\hline    
$\epsilon_{55}$ (eV) & -12.327 & -12.257 & -11.885 & -11.774 & -11.522 \\
$\epsilon_{55}*$ (eV) & -12.326 & -12.254 & -11.840 & -11.772 & -11.483 \\
\hline 
$<(\Delta\epsilon)^2>$ meV$^2$ & 6868 & 4282 & 9 & 1009 & 3627 \\ 
\end{tabular}
\label{table:spindimer}
\end{table}

Applying Eqn. \ref{dft_equation} and normalizing to J$_{4}$, we summarize the following for the values of J$_{1-5}$,

\begin{eqnarray}
J_{1}=6.8\nonumber \\
J_{2}=4.2\nonumber \\
J_{3}=0\\
J_{4}=1\nonumber \\
J_{5}=3.6\nonumber
\label{J_values}
\end{eqnarray}

\noindent We now investigate the low-temperature spin-waves which are characterized by the exchange constants and use these values to motivate a spin-wave model to describe the dispersion curves.

\section{Low temperature spin-waves}

	In this section we discuss the spin-waves measured in the Neel ordered state at low temperatures.  We first present the data measured both along (00L) and along (H00) directions.  We then present a heuristic model for the spin-interactions which map onto an XY Hamiltonian consistent with our analysis of the critical fluctuations which is strongly suggestive of a such a character for the spins.  This model provides a means of estimating the exchange constants along [00L] (J$_{i=3-5}$).  We then derive estimates for the exchange constants within the $a-b$ plane (J$_{i=1,2}$).  Using the values for the exchange constants derived from the spin-waves,  we derive expressions for the  the ordering wave vector and the Curie-Weiss constant and compare with experiment.   We focus our discussion on the spin dynamics along the $c^{*}$ direction as these have the most relevance for the incommensurability.  

\subsection{Experimental Data}

	In this section we present measurements of the spin-waves which characterize the low-temperature magnetic ground state.  To obtain an overview of the magnetic scattering we first conducted preliminary measurements on MARI.  The low temperature spectrum is illustrated in Fig. \ref{mari_inelastic_figure} $a)$ which illustrates a total spin band width of about 4-5 meV and a mode extending to the lowest energies at Q$\sim$ 0.85 \AA$^{-1}$.  This wave vector is consistent with value of magnetic wave vector of $\vec{Q}_{0}$=(1, 0,$\sim$1/7) derived from magnetic structure.

\begin{figure}[t]
\includegraphics[width=95mm]{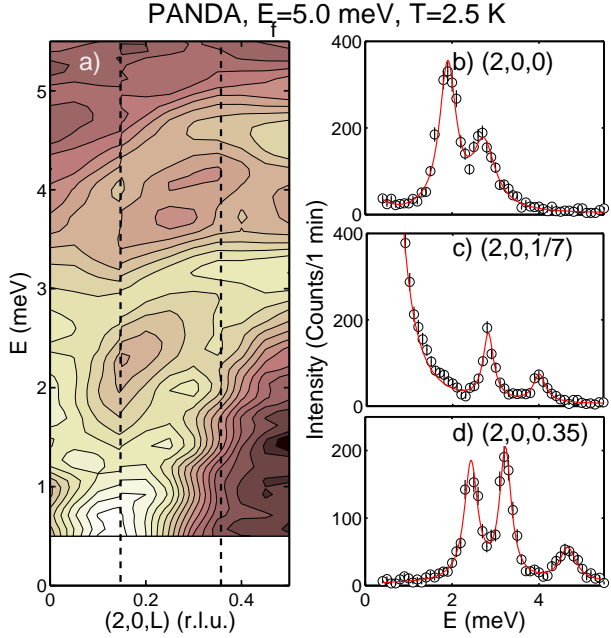}
\caption{(Color online) $a)$ Contour plot of the spin-waves measured along the (00L) direction.  The data are plotted on a logarithmic scale.  Panels $b)-d)$ are representative constant-$Q$ cuts at a serious of wave vectors throughout the Brillouin zone.} 
\label{data_L}
\end{figure}

	To investigate the spin-waves further, we performed measurements on a single crystal using the PANDA cold triple-axis spectrometer.  Given the spiral pitch along the (00L) direction, we focused our measurements on the spin-wave dispersion along $c^{*}$ in an effort to determine the origin of the incommensurate wave vector.  The results are illustrated in Fig. \ref{data_L} where panel $a)$ illustrates a contour plot of the measured intensity and panels $b)-d)$ are representative constant-$Q$ scans.  While the spectrum is quite complex, it can be understood in terms of three branches.  There is one branch which originates from the incommensurate Bragg peak ($\vec{Q}$=(2,0,1/7)) and rises up to $\sim$ 2 meV at the nuclear zone centre ($\vec{Q}$=(200)).  At $\vec{Q}$=(200) we observe two gapped modes.  One disperses up to $\sim$ 4.5 meV at the zone boundary ($\vec{Q}$=(2,0,0.5)) and the other one disperses only up to $\sim$ 3 meV.  The solid curves in Fig. \ref{data_L} $b)-d)$ are fits to Lorentzians without resolution convolution to extract the energy position.

\begin{figure}[t]
\includegraphics[width=70mm]{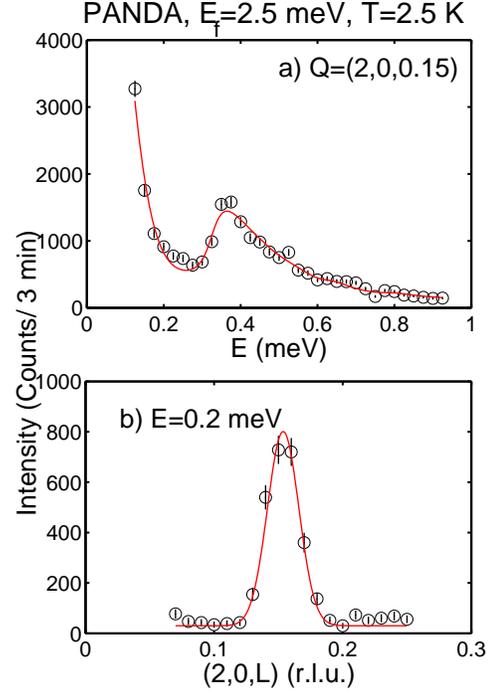}
\caption{$a)$ Constant $Q$ scan at $\vec{Q}$=(2,0,0.15) at T=2.5 K and E$_{f}$=2.5 meV.  A gapped mode is clearly present at 0.35 meV.  Panel $b)$ displays a constant energy scan performed below the gapped mode illustrating the presence of magnetic scattering below the gapped mode and the presence of a gapless phason mode.} 
\label{data_Ef_2p5meV}
\end{figure}

	We investigated the low-energy spin-waves near the incommensurate ordering wave vector of $\vec{Q}$=(2,0,1/7) by using a fixed final energy of E$_{f}$=2.5 meV.  As outlined above in the experimental section, this provided increased energy resolution.  A summary of the results are displayed in Fig. \ref{data_Ef_2p5meV} with panel $a)$ illustrating a constant $Q$ scan performed at the ordering wave vector.  A constant energy scan at low energy transfers of E=0.2 meV is shown in panel $b)$.   The solid curves are resolution convolved fits to a gapped mode and a gapless excitation originating from the incommensurate Bragg peak.   The inclusion of resolution in the fits is necessary to achieve the high-energy tail of the peak in panel $a)$.  The results illustrate the existence of two modes originating from the ordering wave vector at $\vec{Q}$=(2,0,1/7) - one gapless and on gapped with an gap value of 0.35 meV. These modes become degenerate above the gap value within experimental resolution.  The dispersion curves along the (00L) direction are sensitive to the exchange constants along the $c$ direction illustrated in Fig. \ref{structure_figure} and labeled as J$_{3,4,5}$.  A model used to derive these parameters and interpret the physical nature of these modes is presented in the next section.  

\begin{figure}[t]
\includegraphics[width=95mm]{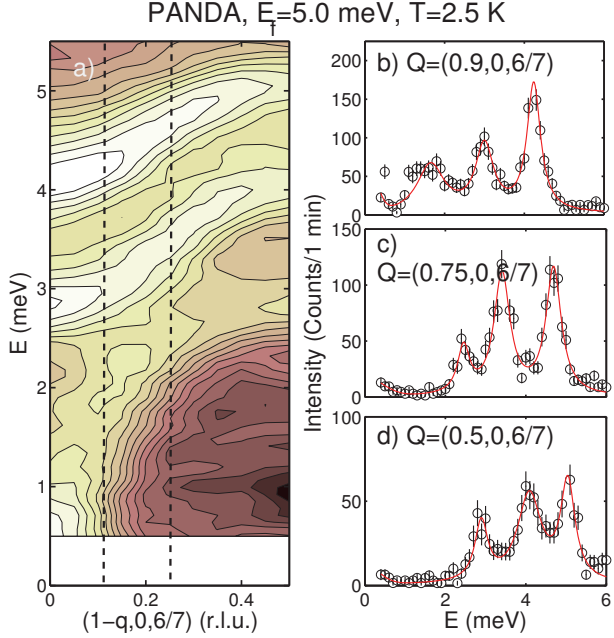}
\caption{(Color online) $a)$ A false contour plot summarizing the constant $Q$ scans along the (1-q,0,6/7) direction taken on PANDA with E$_{f}$=5.0 meV.  Representative constant-$Q$ scans are illustrated in panels $b-d)$.} 
\label{data_a}
\end{figure}

	Having shown the experimental results for the spin-waves along (00L) direction, we now show the results along the (H00) direction.  A summary of a series of constant-$Q$ scans taken along the (H,0,6/7) direction is displayed in Fig. \ref{data_a} with a false contour plot in panel $a)$ and constant $Q$ scans in $b-d)$.  The scans illustrate that the three modes observed at the incommensurate order wave vector disperse and become more well separated along the (H,0,6/7) direction.  The dispersion of these modes are sensitive to the in plane exchange constants, namely the exchange within a given triangle (J$_{1}$) and the exchange between triangles (J$_{2}$).  Values of these exchange constants will be estimated and are presented in the next section.

\subsection{Spin waves along (00L)-XY Hamiltonian}

	Most spiral magnets have the chirality imposed by the Dzyaloshinski-Moriya interaction (characterized by the vector $\vec{D}$) which appears in the Hamiltonian as a cross product of spins ($\vec{D}\cdot\vec{S}_{1}\times\vec{S}_{2}$).  While such a model may fit the low-energy spin waves described above, we find to consistently describe the observed spin-wave band width (defined by $J$) and the incommensurate wave vector (described by the spiral pitch $\alpha=\arctan(D/J)$) would require $D \sim J$.  This is rather unphysical given that $D$ is a relativistic correction related to spin-orbit coupling and hence expected to be small in comparison to the spin-exchange $J$ for a $d^5$ high-spin electronic state with quenched total orbital momentum ($L$=0). Only very small orbital momentum could be expected due to the slightly distorted tetrahedral configuration of Fe$^{3+}$O$_4$, with a slight off-centering of Fe. 

	We have taken a different approach to this problem motivated by the structure (Fig. \ref{structure_figure}), the spin dimer calculations, and the polarized neutron scattering results presented earlier.  We now investigate whether the physical properties, including the incommensurate order wave vector, the spin-waves, the Curie-Weiss constant ($\Theta_{CW}$), and most importantly, the magnetic chirality, can be understood in terms of a Heisenberg only framework.  The Fe$^{3+}$-O$^{2-}$ coordination is plotted in Fig. \ref{structure_figure} and illustrates that all of the exchange paths are super-super exchange involving two oxygen ions.  The structure along the $c$-axis (Fig. \ref{structure_figure} $b)$) illustrates that the two next nearest neighbor interactions (J$_{3}$ and J$_{5}$) have very different exchange paths imposed by the crystal symmetry. We now discuss a model for the spin-waves along the $c$ direction.

	To model the spin-waves, we start with the case of chains of coupled triangles (Fig. \ref{structure_figure} $b)$) and therefore the Hamiltonian for isolated triangles weakly coupled along the $c$-axis.  Experimentally, we believe this is a good starting point owing to the fact that the spiral propagation wave vector is along $c$ and also the fact that the experimental dispersion within the $a-b$ plane are substantially less complicated in comparison to the spin-wave dispersion along the [001] direction as illustrated in comparing Figs. \ref{data_L} and \ref{data_a}.  This approach is also corroborated by the paramagnetic diffuse scattering presented earlier which illustrated that the spin correlations along [001] were significantly stronger than [100].   Based on the coordination paths illustrated in Fig. \ref{structure_figure} we pursue a model based on one nearest neighbor exchange (J$_{4}$) and two next nearest neighbor (J$_{3,5}$).  We therefore write the Heisenberg component of the spin Hamiltonian as,

\begin{eqnarray}
\label{heis_ham}
H_{H,i}=  J_{4}\vec{S}_{i}^{a} \cdot \vec{S}_{i+1}^{a} + J_{3}\vec{S}_{i}^{a} \cdot \vec{S}_{i+1}^{b}+J_{5}\vec{S}_{i}^{a} \cdot \vec{S}_{i+1}^{c}.
\end{eqnarray}

\noindent  The index $i$ correspond to different triangles along the $c$-axis and the $a$, $b$,and $c$ describe the three different spins within a given isolated triangle.  The interactions $J_{3,4,5}$ define the three different exchange pathways between two different triangles along a chain discussed above.  The coupling between $a$ and $a$ spins on different triangles represent the nearest neighbor interaction ($J_{4}$) and $b,c$ spins correspond to next nearest neighbors ($J_{3,5}$).

	To obtain a dispersion for linear spin-waves, we have imposed a ground state defined by the standard 120$^{\circ}$ structure within a triangle and a spiral along $c$, defined by the pitch angle $\alpha$.  Following previous spin-wave calculations for triangular antiferromagnets (Ref. \onlinecite{Jolioceur89:40}), the transformations are represented by the following for first, $a$ spins,

\begin{eqnarray}
S_{j}^{ax}&&=s_{j}^{x} \cos(j \alpha) - s_{j}^{y} \sin(j \alpha) \nonumber \\ 
S_{j}^{ay}&&=s_{j}^{y} \cos(j \alpha) + s_{j}^{x} \sin(j \alpha) \nonumber \\
S_{j}^{az}&&= s_{j}^{z} \nonumber
\end{eqnarray}

\noindent for $b$ spins, 

\begin{eqnarray}
\label{ground_state}
S_{j}^{bx}&&=s_{j}^{x} \cos(j \alpha+2\pi/3) - s_{j}^{y} \sin(j \alpha+2\pi/3) \nonumber \\ 
S_{j}^{by}&&=s_{j}^{y} \cos(j \alpha+2\pi/3) + s_{j}^{x} \sin(j \alpha+2\pi/3) \nonumber \\
S_{j}^{bz}&&= s_{j}^{z}
\end{eqnarray}

\noindent and for $c$ spins,
 
\begin{eqnarray}
\label{ground_state}
S_{j}^{cx}&&=s_{j}^{x} \cos(j \alpha-2\pi/3) - s_{j}^{y} \sin(j \alpha-2\pi/3) \nonumber \\ 
S_{j}^{cy}&&=s_{j}^{y} \cos(j \alpha-2\pi/3) + s_{j}^{x} \sin(j \alpha-2\pi/3) \nonumber \\
S_{j}^{cz}&&=s_{j}^{z}. \nonumber
\end{eqnarray}

\noindent By symmetrizing the Hamiltonian and writing in terms of the new transformed coordinates $s{_i}$, the Heisenberg component of the Hamiltonian takes the following form,

\begin{eqnarray}
\label{heis_ham_trans}
H_{H,n}=\Gamma \left(s_{n}^{x}s_{n+1}^{x}+s_{n}^{y}s_{n+1}^{y}\right) + \Theta \left( s_{n}^{z}s_{n+1}^{z} \right)
\end{eqnarray}

\noindent with

\begin{eqnarray}
\label{sym_def}
\Gamma&&=J_{4}\cos(\alpha)+J_{3}\cos(\alpha+2\pi/3)+J_{5}\cos(\alpha-2\pi/3) \nonumber\\
\Theta&&=J_{3}+J_{4}+J_{5}.
\end{eqnarray}

\noindent This model therefore maps directly onto a Hamiltonian with XY symmetry.  We note that this approach does not directly provide information on the exchange between $S=5/2$ spins in a triangle which is expected to be the strongest interaction.  However, the triangle-triangle coupling described by our model is crucial to understand the spin-waves and chirality along the $c$-axis.
 
	This approach follows closely the analysis applied to Ba$_{2}$CuGe$_{2}$O$_{7}$.~\cite{Zheludev99:59,Zheludev96:54}  The  main difference in our calculation is that we have imposed a chirality to the Hamiltonian through two different next nearest exchange constants J$_{3,5}$ instead of using a Dzyaloshinskii-Moriya interaction as required in Ba$_{2}$CuGe$_{2}$O$_{7}$.

\begin{figure}[t]
\includegraphics[width=90mm]{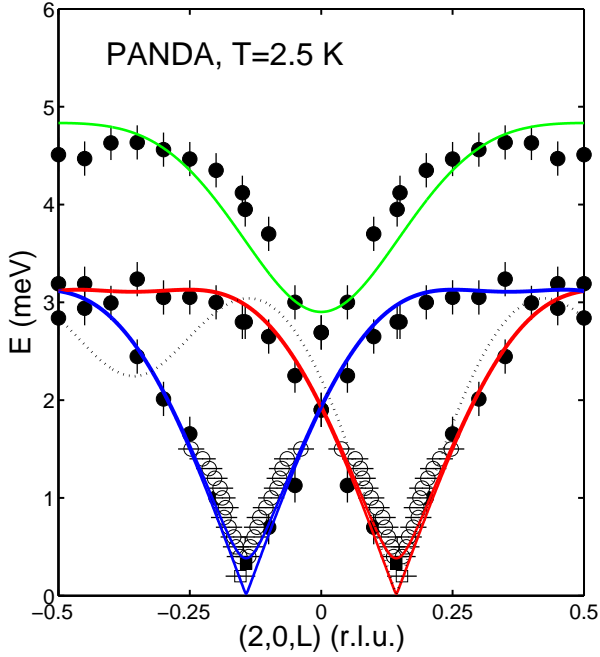}
\caption{ (Color online) A plot of the peak position in energy as a function of momentum transfer along the $c^{*}$ direction (00L).  The solid lines are fits to the spin-wave models discussed in the text.  The dashed line is a calculation with J$_{3}$ fixed to zero and just fitting J$_{4}$ and J$_{5}$.  The data has been symmetrized around L=0 for clarity.} 
\label{dispersion_L}
\end{figure}

	Based on this analysis, we can then use the formulae presented in Ref. \onlinecite{Lovesey_book} to obtain the spin-wave dispersion along (00$\tilde{L}$).  We write the following general formula for the predicted modes.

\begin{eqnarray}
\label{E_q}
E(\tilde{L})=2S\sqrt{A(\tilde{L})^{2}-B(\tilde{L})^{2}}, 
\end{eqnarray}

\noindent with $A_{\tilde{L}}$ and $B_{\tilde{L}}$ are derived as

\begin{eqnarray}
\label{A_B}
A(\tilde{L})&&= \Theta+{(\Gamma-\Theta) \over 2} \cos(2 \pi \tilde{L}) +D_{i} \nonumber\\
B(\tilde{L})&&= \Theta \cos(2 \pi \tilde{L}) - {(\Gamma-\Theta) \over 2} \cos(2 \pi \tilde{L}),
\end{eqnarray}

\noindent with $\Theta$ and $\Gamma$ defined in Eqn. \ref{sym_def}.  We have included a single-ion anisotropy term $D_{i}$ in a similar manner to that used for TbMnO$_{3}$ and LaMnO$_{3}$ to account for gaps in the excitation spectrum, particularly at the ordering wave vector and at the nuclear zone centre.~\cite{Moussa96:54,Senff08:20}  We expect three modes for a given chiral domain.  The first is gapless and associated with rotations of the spin within the spiral plane.  The other two modes correspond to transverse excitations of the spiral plane and may be gapped due to anisotropy.~\cite{Coldea03:68,Senff08:20}

	This model provides a physical understanding of the spin-waves and a fit is illustrated in Fig. \ref{dispersion_L}.  Fig. \ref{dispersion_L} displays three branches-one which is gapped and centered at the nuclear zone center $\vec{Q}$=(2,0,0) (colored green) and two branches centered around the incommensurate positions (colored red and blue).  In our formulation above in terms of $A(\tilde{L})$ and $B(\tilde{L})$ these would correspond to branches such that $\tilde{L}=L$ and $\tilde{L}=L \pm \alpha$.  The solid circles in Fig. \ref{dispersion_L} are obtained from constant-$Q$ scans and the open circles are obtained from constant energy scans done with a fixed E$_{f}$=5.0 meV.  This picture illustrates that the complicated dispersion measured and plotted in Fig. \ref{data_L} is simplified by the fact that one branch is associated with the spin-waves originating from the other incommensurate magnetic Bragg peak.

	The branches near the incommensurate points ($\tilde{L}=L \pm \alpha$) consist of both a gapped and gapless excitations.  The gapless mode is the required phason for a spiral structure and corresponds to a rigid rotation of the spins within the spiral plane.  This mode is reproduced from our spin-wave analysis with $D=0$.  The gapped mode located at the incommensurate position can be interpreted as the result of fluctuations of the spiral plane which follow the symmetry of the magnetic order.  This necessitates a finite $D_{1}$. The large gapped mode at the zone centre ($\tilde{L}=L$) can be interpreted as fluctuations of the spiral plane along the $c$-axis or perpendicular to the spiral plane and can be reproduced with a significant value of $D_{2}$.  These fluctuations have the symmetry of the lattice and not the magnetic order and therefore appear centered around the commensurate positions.  Our analysis here relies on the existence of two anisotropies associated with the gaps at the incommensurate positions and the nuclear zone centre.   

	A fit to the above model is illustrated in Fig. \ref{dispersion_L} (red, blue, and green curves) resulting in the following parameters

\begin{eqnarray}
\label{heis_123}
J_{4}= 0.10 \pm 0.02\ meV \nonumber \\
J_{5}= 0.33 \pm 0.02\ meV \\
J_{3}= 0.13 \pm 0.02\ meV. \nonumber
\end{eqnarray}

\nonumber We also extract two single-ion type anisotropies from the curves associated with in-plane and out of plane fluctuations $D_{1}=0.0040 \pm 0.0001$ meV and $D_{2}=0.31 \pm 0.05$ meV respectively.  These anisotropies result in a finite gap for spin fluctuations.  The large anisotropy found for out of plane fluctuations confirms the strongly XY nature of the magnetic properties as derived from the temperature dependence of the critical spin fluctuations and as postulated based on the magnetic structure.  

	These parameters are broadly consistent with the spin dimer calculations which predict a helical set of exchange constants.  The magnitudes are generally consistent with the calculation except the value of J$_{3}$ is much larger than predicted but is required to provide a good fit to the spin-wave dispersion near the Brillouin zone boundary.  The spin-wave formulation for the excitations along (00L) are quite sensitive to the values of J$_{3}$, particularly near the zone boundary. This is demonstrated in Fig. \ref{dispersion_L} by the dashed line which is a calculation with J$_{3}$ fixed to be zero and J$_{5}$=0.5 meV.  To obtain the relatively flat dispersion of the modes near the zone boundary, a relatively large value of J$_{3}$ is required within our formulation.  The spin-wave calculation is equally sensitive to J$_{4}$.  While we can reproduce the flat nature near the zone boundary and the band-width with competing J$_{4}$ and J$_{5}$ values only, such a model does not fit the initial slope of the modes near the magnetic Bragg peaks.  This spin-wave model therefore provides a sensitive determination of J$_{3,4,5}$.  As a check of our analysis, we now use the derived values of J$_{3,4,5}$ to calculate the ordering wave vector and compare the value with experiment.

\subsubsection{Incommensurate ordering wave vector $q_{0}$}

	The ordering wave vector defined by the pitch $\alpha$ can be obtained by calculating the classical minimum of the Hamiltonian defined above.  The classic exchange energy for the spiral magnetic structure proposed by Marty \textit{et al.} (Ref. \onlinecite{Marty08:101}), independent of any Dzyaloshinskii-Moriya interaction, can be written: 

\begin{eqnarray}
\label{classical_energy}
E(\alpha)= \sum_{i} J_{4}\vec{S}_{i}^{a} \cdot \vec{S}_{i+1}^{a} + J_{3}\vec{S}_{i}^{a} \cdot \vec{S}_{i+1}^{b}+J_{5}\vec{S}_{i}^{a} \cdot \vec{S}_{i+1}^{c}=\nonumber \\
N (J_{4}\cos(\alpha)+J_{5}\cos(\alpha+2\pi/3)+J_{3}\cos(\alpha-2\pi/3)), \nonumber \\
\end{eqnarray}

\noindent where N is the number of cells considered along the c-axis, and the sum (\textit{i}) runs from 1 to \textit{N}. This is the energy for the left-handed crystal structure, and the negative triangular chirality. The energy for the positive triangular chirality can be obtained by changing 2$\pi$/3 into -2$\pi$/3 and vice-versa. The in-plane exchange energy is here irrelevant because it does not depend on $\alpha$. The magnetic modulation is considered incommensurate ($\tau\sim\frac{1}{7}$), i.e. 
the summation runs over an infinite number of sites and only normal terms contribute to the energy (Umklapp terms sum-up to zero for an incommensurate structure).
The resulting functional of the energy is extremely interesting, because it is not invariant by a change of sign of $\alpha$. Changing $\alpha$ to -$\alpha$ is equivalent to permuting $J_3$ and $J_5$ and our previous analysis 
showed that these two exchange interaction have very different strengths. This demonstrates that for a given triangular chirality, a single magnetic chirality is stabilized by isotropic exchanges alone, whose strengths are imposed by the chiral crystal structure.  
\noindent The stability condition for the energy ($dE/d\alpha$=0) leads to the following incommensurability (defined by $\tan(\alpha)$) in terms of the three exchange parameters ($J_{3,4,5}$) defined above:

\begin{eqnarray}
\label{classical_min}
\tan(\alpha)={{\sqrt{3} (J_{3}-J_{5})} \over {2J_{4}-(J_{5}+J_{3})}}.
\end{eqnarray}
  Substituting the values extracted from the spin-wave dispersion along $c^{*}$ for J$_{3,4,5}$, we find 
\begin{eqnarray}
\label{q0_exp}
q_{0}=\frac{\alpha}{2\pi}=0.148 \pm 0.007\ r.l.u.,
\end{eqnarray}
\noindent in very good agreement with the experimental value of 0.143 r.l.u. For the opposite triangular chirality, the value of q$_{0}$ will be reversed. Both structures are compatible with the results of Marty \textit{et al.} derived from neutron polarimetry \cite{Marty08:101}. This illustrates a consistent description of the incommensurate wave vector and the spin-waves in terms of a Heisenberg only model for the spin interactions.  This result also corroborates the spin-wave analysis which implies that all three competing exchange constants are required to describe the dynamics.  We note that a J$_{4}$-J$_{3}$ or a J$_{4}$-J$_{5}$ only picture cannot describe the incommensurate wave vector.  We emphasize also that while the Dzyaloshinskii-Moriya interaction is present owing to the low crystal symmetry, our analysis shows that it is not required and would only be a small perturbation to the energy \emph{asymmetry} between J$_3$ and J$_5$ which determines the ordering wave vector. In fact, one can rewrite equation \ref{classical_energy} to include three different Dzyaloshinskii-Moriya interactions alongside the three exchange paths.  We consider only the z-component of the DM vectors, that couples to the in plane spin structure, labeled by DM$_i$ (i=3,4,5). The stability condition becomes:
\begin{eqnarray}
\label{classical_min}
\tan(\alpha)={{\sqrt{3} (J_{3}-J_{5})+2DM_4+DM_3-DM_5} \over {2J_{4}-(J_{5}+J_{3})+\sqrt{3}(DM_5-DM_3)}}.
\end{eqnarray}
 It is clear that for Fe$^{3+}$, the strengths of the DM terms (proportional to $\frac{\Delta g}{g}J$) will be at best only of few percents of the isotropic terms and will not play any role in selecting a given magnetic chirality, unless for accidental degeneracies of the different J terms. The values extracted from the spin-wave analysis indicate that this is not the case here.
 
	Since the single magnetic chirality (sign of $\alpha$) depends on the initial choice of the triangular chirality, the question remains about the absolute chirality of the system. However, it is important to realize that the magnetic structure corresponding to the positive and negative triangular chiralities are supported by different irreducibles representations of the P321 group (totally symmetric $\tau_1$ for positive triangular chirality and $\tau_2$ for negative). The presence of single ion anisotropy terms, which are present (however small) due to the two-fold symmetry on the Fe$^{3+}$ site, will actually select a unique triangular chirality, a unique magnetic state will be stabilized. 

\subsubsection{Spin waves within the $a-b$ plane}

\begin{figure}[t]
\includegraphics[width=90mm]{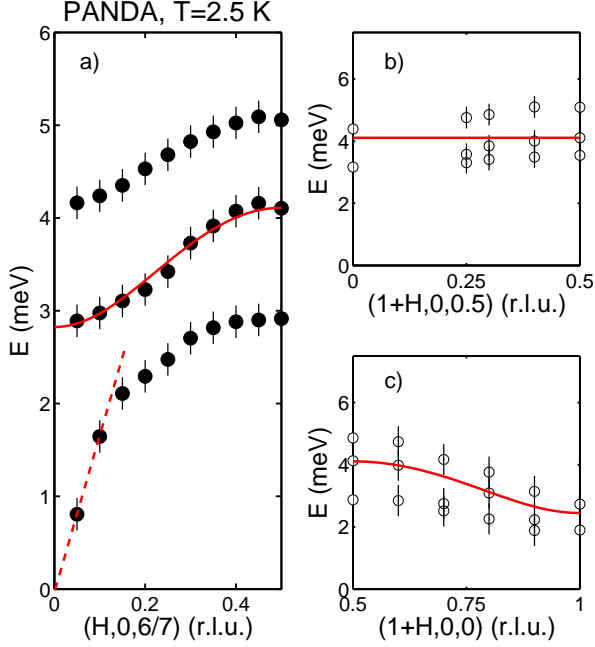}
\caption{(Color online) The dispersion curves within the $ab$ plane extracted from the analysis described in the text.  $a)$ illustrates the peak positions along the (H,0,6/7) direction.  The dashed line is used to estimate the exchange within a triangle (J$_{0}$).  The solid lines derived from the general expression described in the text for antiferromagnetic exchange.  $b)$ plots the peak positions around the zone boundary and illustrates relatively little dispersion, indicating small next-nearest neighbor exchange.  $c)$ shows the dispersion along the (1+H,0,0) direction.  The solid curve is a calculation to the same formula used in panel $a)$, but with a finite value of $\Delta$ to account for not being exactly at the magnetic zone center. } 
\label{dispersion_ab}
\end{figure}

	To extract information on the exchange parameters within the $a-b$ plane, we investigated the spin-wave dispersion along the (H,0,$6/7$) and (H,0,0) directions and along the zone boundary (H,0,0.5). The spin-wave dispersion along (H,0,$6/7$) is presented in Fig. \ref{dispersion_ab} $a)$.  While our analysis above for the spin-waves along the (0,0,L) direction do not provide direct information on the exchange within a triangle J$_{1}$, the initial slope of the lowest energy mode along the (H,0,$1/7$) can be interpreted as a first approximation for this exchange parameter.  Using the second moment sum rule (Ref. \onlinecite{Hohenberg}) which relates S($\vec{Q}$,E) to the dispersion $E(\vec{Q})$ as $\langle E(\vec{Q})^{2}\rangle=\int dE E^{2} S(\vec{Q},E)/\int dE S(\vec{Q},E)$,

\begin{eqnarray}
\langle E(\vec{Q})^{2}\rangle={2\over 3}S(S+1)\sum_{\vec{d}}J_{\vec{d}}^{2} [1-\cos(\vec{Q}\cdot\vec{d})],
\end{eqnarray}

\noindent and assuming small momentum transfers $q$ close to the magnetic Bragg peak and parallel to [100], we write,

\begin{eqnarray}
\lim_{q \to 0} E(q)^{2}=S (S+1) \delta^{2} J_{1}^{2} q^{2} = (\hbar c q)^{2},
\end{eqnarray}

\noindent with $\delta$ being the nearest neighbor distance.  Substituting a spin-wave velocity of 17.5 meV \AA\ derived from the dashed line in Fig. \ref{dispersion_ab} $a)$ and a bond length $\delta \sim$ 3.7 \AA (Table \ref{table:spindimer}), we obtain J$_{1}$=1.6$\pm$ 0.3 meV.   

	To derive values for the exchange constant between the triangles within the $ab$ plane, we consider couplings (J$_{2}$ as indicated in Fig. \ref{structure_figure}) between triangles and use the following expression for an antiferromagnet,

\begin{eqnarray}
\label{q0_exp}
E(H)^{2}=4S^{2}\left(\Delta^2+J_{2}^{2}\sin^{2}(2\pi H))\right).
\end{eqnarray}

\noindent Here we have inserted an extra parameter $\Delta$ to take into account that we are not exactly at the zone magnetic zone center for dispersion curves like those shown in Fig. \ref{dispersion_ab} $c)$ or for the presence of anisotropies as discussed above in regards to the spin-waves along the (00L) direction.   The solid curves in Fig. \ref{dispersion_ab} $a)$ and $c)$ are calculations with the above formula with J$_{2}$=0.45 meV.  

	The spin-wave dispersion around the zone boundary is illustrated in Fig. \ref{dispersion_ab} $b)$ and is sensitive to higher order (beyond nearest neighbor) interactions between the triangles.  While it is difficult to separate the modes along this direction, our results do show that there is relatively little dispersion along this direction in comparison to the energy values.  This indicates a negligible next nearest neighbor interaction between the triangles and such a scenario is represented by the flat line in Fig. \ref{dispersion_ab} $b)$.  

	In summary, based on the spin-wave dispersion curves in the $a-b$ plane, we extract the following for the exchange interactions

\begin{eqnarray}
\label{heis_a}
J_{1}= 1.6 \pm 0.3\ meV \nonumber \\
J_{2}= 0.35 \pm 0.05\ meV. \\
\end{eqnarray}

\noindent We now compare the exchange parameters obtained to the Curie-Weiss temperature previously reported from magnetization measurements.

\section{Curie-Weiss Temperature}

	Based on our values for the exchange constants, we can obtain an estimate for the Curie-Weiss temperature ($\Theta_{CW}$) using the following formula. 

\begin{eqnarray}
\label{CW_law}
k_{B}\Theta_{CW}={1\over 3} S(S+1) \sum\limits_{n}J_{n},
\end{eqnarray}

\noindent where the sum is performed over nearest neighbors.  The above result is derived from a mean-field approach and is based on the molecular field produced at a given site by the nearest neighbors and where the Hamiltonian can be reduced to a form $\sum\limits_{n} \vec{S}_{n} \cdot \vec{H}_{W}$, where $\vec{H}_{W}$ is the molecular or Weiss field.~\cite{Goodenough_book}  Owing to the fact that the magnetic structure in Ba$_{3}$NbFe$_{3}$Si$_{2}$O$_{14}$ is quite complex, it is not entirely clear how the Hamiltonian can be simplified to such a form to provide a direct value for the Curie-Weiss constant.  Nevertheless, we can estimate the value as follows based on the spin-wave analysis described above and the number of nearest neighbors for a given Fe$^{3+}$ ion,

\begin{eqnarray}
\label{CW_law}
\Theta_{CW}= {1\over 3} S(S+1)...\nonumber\\
 2\times(J_{1}+J_{2}+J_{3}+J_{4}+J_{5})/k_{B}\nonumber \\
\sim 170 K.
\end{eqnarray}

\noindent The terms for $J_{3,4,5}$ are quite rigorous as there are two nearest neighbors.  We have chosen to count $J_{1}$ twice and $J_{2}$ also 2 times as each Fe$^{3+}$ only has two neighbors coupled by each exchange pathway.  The value obtained above is in reasonable agreement with the measured value of $\Theta_{CW}$=-173 K and -190 K described in Refs. \onlinecite{Marty10:81,Marty08:101,Zhou09:102}.  The comparison confirms the validity of our spin-wave analysis and the extracted parameters.  It also confirms our assertion that a helical Heisenberg model is more appropriate to describe the physical properties Ba$_{3}$NbFe$_{3}$Si$_{2}$O$_{14}$ rather than a model invoking a large Dzyaloshinskii-Moriya interaction term in the Hamiltonian.

\section{Conclusions}

	We have measured the spin fluctuations in Ba$_{3}$NbFe$_{3}$Si$_{2}$O$_{14}$ at both low-temperatures and around T$_{N}$ with the goal of understanding the ordering wave vector and the critical properties.  We have found the critical dynamics are consistent with those of other two dimensional triangular magnets with a 120$^{\circ}$ magnetic structure.  However, the critical wave vector and chirality are independent of temperature and a chiral nature of the diffuse scattering remains at temperatures well above T$_{N}$.  Through an analysis of the structure using a spin dimer calculation, we have followed a scheme where the exchange constants along the $c$-axis are helical with two different next nearest neighbor couplings.   Using this and the fact that an XY picture seems appropriate we derived the spin-waves for modes propagating along the $c$-axis with the spins confined to the $a-b$ plane and have derived the exchange constants.    The experimental values are reasonable agreement with the spin dimer calculation and reproduce the correct magnetic ordering wave vector and Curie-Weiss constants within error.  These results indicate that the structural and magnetic chiralities are strongly coupled in Ba$_{3}$NbFe$_{3}$Si$_{2}$O$_{14}$ and that symmetric Heisenberg exchange is required for this.   These results point to a unique magnetic system where the chirality is imposed by the nuclear structure and not the Dzyaloshinskii-Moriya interaction.

\begin{acknowledgements}

We are grateful to A. Orszulik (ISIS) and H. Schneider (JCNS) for expert technical support throughout the experiments and to R. Coldea and R. Cowley for helpful discussions.  We are also  grateful for grants from EU-NMI3 Access Program for partial funding of this research. The work at Rutgers was supported by the DOE under Grant No. DE-FG02-07ER46382.

\end{acknowledgements}




\thebibliography{}

\bibitem{Collins97:75} M.F. Collins and O.A. Petrenko, Can. J. Phys. {\bf{75}}, 605 (1997).
\bibitem{Ramirez01:13} A.P. Ramirez, in Handbook of Magnetic Materials, K.J.H. Buschow Ed. (Elsevier Science, Amsterdam, 2001) vol. 13, pp. 423-520.
\bibitem{Kawamura01:79} H. Kawamura, Can. J. Phys. {\bf{79}}, 1447 (2001).
\bibitem{Kawamura98:10} H. Kawamura, J. Phys. Condens. Matter {\bf{10}}, 4707 (1998).
\bibitem{Anderson73:8} P.W. Anderson, Mater. Res. Bull. {\bf{8}}, 153 (1973).
\bibitem{Coldea03:68} R. Coldea, D.A. Tennant, and Z. Tylczynski, Phys. Rev. B {\bf{68}}, 134424 (2003).
\bibitem{Nakatsuji05:309} S. Nakatsuji, Y. Nambu, H. Tonomura, O. Sakai, S. Jonas, C. Broholm, H. Tsunetsugu, Y. Qiu, Y. Maeno Science {\bf{309}}, 1697 (2005).
\bibitem{Shimizu03:91} Y. Shimizu, K. Miyagawa, K. Kanoda, M. Maesato, and G. Saito, Phys. Rev. Lett. {\bf{91}}, 107001 (2003).
\bibitem{Helton07:98} J.S. Helton, K. Matan, M.P. Shores, E.A. Hytko, B.M. Bartlett, Y. Yoshida, Y. Takano, A. Suslov, Y. Qiu, J.-H. Chung, D.G. Nocera, and Y.S. Lee, Phys. Rev. Lett. {\bf{98}}, 107204 (2007).
\bibitem{Lee07:6} S.-H. Lee, H. Kikuchi, Y. Qiu, B. Lake, Q. Huang, K. Habicht, and K. Kiefer, Nature Materials {\bf{6}}, 853 (2007).
\bibitem{Stock09:103} C. Stock, L.C. Chapon, O. Adamopoulos, A. Lappas, M. Giot, J.W. Taylor, M.A. Green, C.M. Brown, and P.G. Radaelli, Phys. Rev. Lett. {\bf{103}}, 077202 (2009). 
\bibitem{Kenzelmann05:95} M. Kenzelmann, A.B. Harris, S. Jonas, C. Broholm, J.Schefer, S.B. Kim, C.L. Zhang, S.-W. Cheong, O.P. Vajk, and J.W. Lynn, Phys. Rev. Lett. {\bf{95}}, 087205 (2005).
\bibitem{Vajk05:94} O.P. Vajk, M. Kenzelmann, J.W. Lynn, S.B. Kim, and S.-W. Cheong, Phys.Rev. Lett/ {\bf{94}}, 087601 (2005).
\bibitem{Senff07:98} D. Senff, P. Link, K. Hradil, A. Hiess, L.P. Regnault, Y. Sidis, N. Aliouane, D.N. Argyriou, and M. Braden, Phys. Rev. Lett. {\bf{98}}, 137206 (2007).
\bibitem{Kenzelmann07:98} M. Kenzelmann, G. Lawes, A.B. Harris, G. Gasparovic, C. Broholm, A.P. Ramirez, G.A. Jorge, M. Jaime, S. Park, Q. Huang, A.Ya. Shapiro, and L.A. Demianets, Phys. Rev. Lett. {\bf{98}}, 267205 (2007).
\bibitem{Kimura03:426} T. Kimura, T. Goto, H. Sintani, K. Ishizaka, T. Arima, and Y. Tokura, Nature {\bf{426}}, 55 (2003).
\bibitem{Kitaura04:69} M. Kitaura, K. Mochizuki, Y. Inabe, M. Itoh, H. Nakagawa, S. Oishi, Phys. Rev. B {\bf{69}}, 115120 (2004).
\bibitem{Zhou09:102} H.D. Zhou, C.R. Wiebe, Y.-J. Jo, L. Balicas, R.R. Urbano, L.L. Lumata, J.S. Brooks, R.L. Kuhns, A.P. Reyes, Y. Qiu, J.R.D. Copley, and J.S. Gardner, Phys. Rev. Lett. {\bf{102}}, 067203 (2009).
\bibitem{Lumata10:81} L. L. Lumata, T Besara, P. L. Kuhns, A. P. Reyes, H. D. Zhou, C. R. Wiebe, L. Balicas, Y. J. Jo, J. S. Brooks, Y. Takano, M. J. Case, Y. Qiu, J. R. D. Copley, J. S. Gardner, K. Y. Choi, N. S. Dalal, and M. J. R. Hoch Phys. Rev. B {\bf{81}}, 224416 (2010).
\bibitem{Marty10:81} K. Marty, P. Bordet, V. Simonet, M. Loire, R. Ballou, C. Darie, J. Kljun, P. Bonville, O. Isnard, P. Lejay, B. Zawilski, and C. Simon, Phys. Rev. B {\bf{81}}, 054416 (2010).
\bibitem{Marty08:101} K. Marty, V. Simonet, E. Ressouche, R. Ballou, P. Lejay, and P. Bordet, Phys. Rev. Lett. {\bf{101}}, 247201 (2008).
\bibitem{Zhou09:21} H.D. Zhou, L.L. Lumata, P.L. Kuhns, A.P. Reyes, E.S. Choi, N.S. Dalel, J. Lu, Y.J. Lo, L. Balicas, J.S. Brooks, and C.R. Wiebe, Chem. Mater. {\bf{21}}, 156 (2009).
\bibitem{Shirane_book} G. Shirane, S. Shapiro, and J.M. Tranquada, \textit{Neutron Scattering with a Triple-Axis Spectrometer} (Cambridge Press, 2002).
\bibitem{Stock04:69} C. Stock, W.J.L. Buyers, R. Liang, D. Peets, Z. Tun, D. Bonn, W.N. Hardy, and R.J. Birgeneau, Phys. Rev. B {\bf{69}}, 014502 (2004).
\bibitem{Christianson02:66} R.J. Christianson, Y.J. Wang, S.C. LaMarra, R.J. Birgeneau, V. Kiryukhin, T. Masuda, I. Tsukada, K. Uchinokura, B. Keimer, Phys. Rev. B {\bf{66}}, 174105 (2002).
\bibitem{Zinkin96:54} M.P. Zinkin, D.F. McMorrow, J.P. Hill, R.A. Cowley, J.-G. Lussier, A. Giban, G. Grubel, and C. Sutter, Phys. Rev. B {\bf{54}}, 3115 (1996).
\bibitem{Halpern39:55} O. Halpern and M.H. Johnson Phys. Rev. {\bf{55}}, 898 (1939).
\bibitem{Blume63:130} M. Blume, Phys. Rev. {\bf{130}}, 1670 (1963).
\bibitem{Moon69:181} A.T. Moon, T. Riste, and W.C. Koehler, Phys. Rev. {\bf{181}}, 920 (1969).
\bibitem{Schweika10:xx} W. Schweika, J. Phys: Conf. Series {\bf{211}}, 012026 (2010).
\bibitem{Shirane83:28} G. Shirane, R. Cowley, C. Majkrzak, J.B. Sokoloff, B. Pagonis, and C.H. Perry, and Y. Ishikawa, Phys. Rev. B {\bf{28}}, 6251 (1983). 
\bibitem{Kawamura88:63} H. Kawamura, J. Appl. Phys. {\bf{63}}, 3086 (1988).
\bibitem{Whangbo05:7}  M.-H. Wangbo, D. Dai, and H.-J. Koo, Solid State Sciences {\bf{7}}, 827 (2005).
\bibitem{Koo06:45} H.-J. Koo, Inorganic Chemistry, {\bf{45}}, 10743 (2006).
\bibitem{Halperin69:188} B. I. Halperin and P. C. Hohenberg, Phys.Rev. B {\bf{188}}, 898 (1969). 
\bibitem{Stock10:xx} C. Stock, S. Jonas, C. Broholm, S. Nakatsuji, Y. Nambu, K. Onuma, Y. Maeno, and J.-H. Chung, unpublished (cond-mat/0059962).
\bibitem{Stock08:77} C. Stock, W.J.L. Buyers, Z. Yamani, Z. Tun, R.J. Birgeneau, R. Liang, D. Bonn, and W.N. Hardy Phys.Rev. B {\bf{77}}, 104513 (2008).
\bibitem{Jolioceur89:40} Th. Jolicoeur and J.C. Le Guillou, Phys. Rev. B {\bf{40}}, 2727 (1989).
\bibitem{Zheludev99:59} A. Zheludev, S. Maslov, G. Shirane, I. Tsukada, T. Masuda, K. Uchinokura, I. Zalizynak, and R. Erwin, Phys. Rev. B {\bf{59}}, 11432 (1999).
\bibitem{Zheludev96:54} A. Zheludev, G. Shirane, Y. Sasago, N. Kiode, and K. Uchinokura, Phys. Rev. B {\bf{54}}, 15163 (1996).
\bibitem{Lovesey_book} S.W. Lovesey, \textit{Theory of Neutron Scattering from Condensed Matter} (Clarendon Press, Oxford, 1984).
\bibitem{Moussa96:54} F. Moussa, M. Hennion, J. Rodriguez-Carvajal, H. Moudden, L. Pinsard, and A. Revcolevschi, Phys. Rev. B {\bf{54}}, 15149 (1996).
\bibitem{Senff08:20} D. Senff, N. Aliouane, D.N. Argyriou, A. Hiess, L.P. Regnault, P. Link, K. Hradil, Y. Sidis, and M. Braden, J. Phys.: Condens. Matter {\bf{20}}, 434212 (2008).
\bibitem{Goodenough_book} J. B. Goodenough, \textit{Magnetism of the Chemical Bond} (Interscience Publishers, London, 1963).
\bibitem{Hohenberg} P. C. Hohenberg and W. F. Brinkman, Phys. Rev. B {\bf 10}, 128 (1974).
\end{document}